# Time-space fabric underlying anomalous diffusion

W. Chen

*Institute of Applied Physics and Computational Mathematics, P.O. Box 8009, Division Box 26, Beijing 100088, China (March 2, 2005)*

**This study unveils the time-space transforms underlying anomalous diffusion process. Based on this finding, we present the two hypotheses concerning the effect of fractal time-space fabric on physical behaviors and accordingly derive fractional quantum relationships between energy and frequency, momentum and wavenumber which further give rise to fractional Schrodinger equation. As an alternative modeling approach to the standard fractional derivatives, we introduce the concept of the Hausdorff derivative underlying the Hausdorff dimensions of metric spacetime. And in terms of the proposed hypotheses, the Hausdorff derivative is used to derive a linear anomalous transport-diffusion equation underlying anomalous diffusion process. Its Green's function solution turns out to be a new type of stretched Gaussian distribution and is compared with that from the Richardson's diffusion equation.**

Anomalous diffusion is one of the most important concepts in modern physics[1-4] and is present in extremely diverse engineering fields such as charges transport in amorphous semiconductor,[5] vibration and acoustic dissipation in soft matter,[6] magnetic plasma,[7] polymer dynamics,[8] turbulence[9] and quantum processes[10] among many other problems.[4] However, it is noted that the anomaly is mostly introduced in a descriptive level of statistical representation or phenomenological modeling. The purpose of this communication is to investigate time-space origin of anomalous diffusion. And some new results as summarized in the abstract are introduced in this study.

## I. FRACTAL TIME-SPACE TRANSFORMS

The definition of anomalous diffusion is based only on the time evolution of the mean square displacement of diffusing particle (random walkers) movements[1-3,11,12]

$$\langle \Delta x^2 \rangle \propto \Delta t^\eta. \tag{1}$$

where $\Delta x$ represents distance, $\Delta t$ denotes time interval, $\eta$ is a positive real number, and the brackets represent the mean value of random variables (e.g., a collection of particles). For anomalous diffusion ($\eta \neq 1$), particles move coherently for long times with infrequent changes of direction, faster in superdiffusion ($\eta \succ 1$) and more slowly in subdiffusion ($\eta \prec 1$) than linearly with time in normal diffusion ($\eta = 1$), i.e., $\langle \Delta x^2 \rangle \propto \Delta t$. The celebrated anomalous diffusion equation of fractional derivatives is typically used in the corresponding phenomenological continuum modeling, which in isotropic linear media is stated as[1,3,4,13]

$$\frac{\partial^\alpha s}{\partial t^\alpha} + \gamma \left( -\nabla^2 \right)^\beta s = 0, \qquad 0 \prec \alpha, \beta \leq 1, \tag{2}$$

where $s$ is the physical quantity of interest (e.g., temperature in anomalous heat conduction), $\gamma$ the corresponding physical coefficient, $\left(-\nabla^2\right)^\beta$ represents the symmetric fractional Laplacian[14], and $\alpha$ and $\beta$ can be real numbers. The fundamental solution of equation (2) is the time-dependent Lévy probability density distribution (fat tailed distribution $\alpha$=1, $\beta$<1), in which $2\beta$ is the stability index of Lévy distribution[1-3]. Equation (2) also underlies the fractional Brownian motion (long time range correlation, $\alpha$<1, $\beta$=1), in which $\alpha$ is the memory strength index of process[4,13], and the smaller $\alpha$, the stronger memory. These two anomalous statistics are often considered the statistical mechanism leading to anomalous diffusion (1) and accordingly $\eta = \alpha/\beta$ can be derived.[1-5,15] When $\alpha$=1, $\beta$<1, equation (1) leads to the diverging moment of higher than $2\beta$ order

$$\langle (\Delta x)^n \rangle = \infty, \qquad 2\beta < n. \tag{3}$$

For $\beta$ <1, the mean square displacement diverges,[1-4,15,16] which implies the potential energy can not trap the particle. Following this view, the mean kinetic energy for a finite mass $m$ also diverges.[16] To solve this paradox, we introduce the following scaling transforms to have the new observation metric spacetime,

$$\begin{cases} \Delta \hat{x} = \Delta x^\beta \\ \Delta \hat{t} = \Delta t^\alpha \end{cases}, \qquad 0 \prec \alpha, \beta \leq 1. \tag{4}$$

The above metric transforms (4) coincide with the classical definition of the Hausdorff time-space dimension.[17] The anomalous diffusion scaling ($\eta = \alpha/\beta \neq 1$) of the mean square displacement (1) is recast as a normal diffusion under the new metric spacetime

$$\langle \Delta \hat{x}^2 \rangle \propto \Delta \hat{t}, \tag{5}$$

where the second moment is finite and the corresponding mean kinetic energy exists. It is worth pointing out that the corresponding definition of velocity needs to be changed [see Eq. 15 further below], and thus the quantity of kinetic energy varies accordingly. (4) and (5) explicitly displays the fractal metric spacetime origin of anomalous diffusion process. Unlike the classical Lorentz transforms in the special relativity, the spacetime transforms (4) are nonlinear in nature and is not concerned with the frame of moving inertial reference.

It is possible to combine the transforms (4) with the Galilean and the Lorenz transforms in which the concept of velocity in the fractal metric spacetime must be redefined by the fractional derivative or the Hausdorff derivative to be defined later on [also see Eq. (15) further below]. The time scaling transform in (4) was also proposed by Hoffmann[18] and Li[3], referred to as "internal clock", to solve counterintuitive paradox on the entropy production of anomalous diffusion process.

In terms of the transforms (4), Lévy statistics and fractional Brownian motion are considered a consequence of the fractal metric spacetime, while the classical Gaussian distribution and Brownian motion correspond to the limiting $\alpha=1$ and $\beta=2$ spacetime fabric, respectively. On the other hand, the restoration of the normal diffusion formalism in (5) implies the invariance of physical law under scale transforms and equivalence between anomalous environmental effect and scale time-space geometry, which is a reminiscent of the two pillar principles of general covariance and equivalence in the general relativity. Generalizing these observations, this study conjectures the following two hypotheses:

1) **The hypothesis of fractal invariance: the laws of physics are invariant regardless of the fractal (scale) metric spacetime (coordinate systems)**.

2) **The hypothesis of fractal equivalence: the influence of anomalous environmental fluctuations on physical behaviors equals that of the time-space transforms (4).**

The first hypothesis means that the general form of physical equations would be invariant under the fractal transformations (4). The second one suggests that the anomaly in physical behaviors (e.g, anomalous diffusion) is caused by environmental effect (field noise) and can fully be explained and represented by the scale spacetime geometry (4). The hypothesis of fractal invariance is very similar to the so-called scale relativity principle pioneered by Nottale.[19] Unlike the latter, this study does not intend to incorporate the Einstein's relativistic effects arising from the reference frame of motion transforms such as acceleration and velocity (inertial). To my best understanding, this study also develops different time-space transforms, calculus, statistics, and physics formalisms and pursues distinct problems compared with Nottale's. The following sections will substantiate the above two heuristic fractal hypotheses and the transforms (4) through the introduction of new mathematical formalisms and typical applications.

## II. FRACTAL SPACETIME ORIGIN OF FRACTIONAL QUANTUM MECHANICS

Time and space are very fundamental concepts in nature and give rise to diverse mathematical theories and physical quantities. Therefore, the time-space transforms (4) will have an impact on general sciences and engineering. Serving as an illustrating example, this section applies the foregoing fractal hypotheses and transforms to quantum mechanics. According to the hypothesis of fractal invariance, the quantum relationships between energy and frequency, momentum and wavenumber in the fractal time-space (4) remain the classical linear formalism

$$E = \hat{h}_\alpha \hat{v}, \tag{6}$$

$$p = \hat{h}_\beta \hat{k}, \tag{7}$$

where $E$ represents energy, $p$ denotes momentum, $\hat{h}_\alpha$ and $\hat{h}_\beta$ are the scaled Planck constant thanks to the scale spacetime, $\hat{k}$ wavenumber and $\hat{v}$ frequency. In terms of the transforms (4), it is straightforward to connect the wavenumber and frequency measures between the two metric spacetime

$$\hat{v} = v^\alpha \quad \text{and} \quad \hat{k} = k^\beta. \tag{8}$$

Thus, we have

$$E = \hat{h}_\alpha v^\alpha, \quad 0 \prec \alpha \leq 1, \tag{9}$$

$$p = \hat{h}_\beta k^\beta, \quad 0 \prec \beta \leq 1. \tag{10}$$

As discussed before, $\alpha$ and $\beta$ are the statistical indices of fractional Brownian motion and Lévy process, respectively. Therefore, the fractional quantum (9) and (10) imply that Lévy statistics and fractional Brownian motion are essentially related to momentum and energy, respectively. The kinetic energy remains $E_k = |p|^2/2m$, where $m$ is the particle mass, whereas $E_k = D_\beta |p|^{2\beta}$ given in refs. 20 and 21 is incorrect, where $D_\beta$ is the scaled constant with the physical dimension erg$^{1-2\beta}$×m$^{2\beta}$×sec$^{-2\beta}$.[20]

Based on the hypothesis of the fractal invariance, a quantum plane wave function is stated as $\Psi(\hat{x}, \hat{t}) = Ae^{i\hat{k}\hat{x} - i\hat{v}\hat{t}}$ in the transformed metric spacetime. By using the fractional quantum relationships (9) and (10), it is easy to construct a quantum Hamiltonian through a plane wave analysis in the classical fashion and then derive the fractional Schrodinger equation

$$e^{i\pi\alpha/2}\hat{h}_\alpha \frac{\partial^\alpha \Psi}{\partial t^\alpha} = \frac{\hat{h}_\beta^2}{2m}(-\Delta)^\beta \Psi + V\Psi, \quad 0 \prec \alpha, \beta \leq 1, \tag{11}$$

where $V$ represents the potential energy. In terms of the hypothesis of the fractal equivalence, the fractional Schrodinger equation (11) accounts for the affect of scale metric spacetime on quantum processes. In the literature,[20-24] the above fractional Schrodinger equation were derived either based on the quantum integral over the Lévy paths in contrast to the conventional Feynman Gaussian path integral[20,21] or the fractional time derivative representation of the fractional Brownian motion.[23,24] Unlike this study, none of these derivations, however, is a consequence of a basic principle of physics. The fractional quantum mechanics has been found useful in modeling complex quantum systems such as polymers[21] and is of great potential use in quantum phenomena in which anomalous diffusion[10,25] and Lévy statistics (e.g., laser cooling[26]) presents prominently.

## III. HAUSDORFF DERIVATIVE, ANOMALOUS DIFFUSION AND STRETECHED GAUSSIAN

In recent decade the fractional derivative has widely been used in the analysis and modeling of anomalous diffusion. As an alternative modeling formalism, this study introduces the concept of the Hausdorff derivative of a function $g(t)$ with respect to a fractal measure $t$

$$\frac{\partial g(t)}{\partial t^\alpha} = \lim_{t' \to t} \frac{g(t)-g(t')}{t^\alpha - t'^\alpha} = \frac{\partial g(\hat{t})}{\partial \hat{t}}. \tag{12}$$

The Hausdorff derivative (12) differs from the standard fractional derivative in that it does not involve the integral convolution and is local in nature. Note that the symbol of the Hausdorff derivative differs from that of the fractional derivative in that index $\alpha$ appears only once. In the same manner, we can also develop the Hausdorff integral formalism. The elementary physical concepts such as velocity in a fractal spacetime $(x^\beta, t^\alpha)$ can be redefined by

$$\hat{v} = \frac{d\hat{x}}{d\hat{t}} = \frac{dx^\beta}{dt^\alpha}, \qquad \hat{t}, \hat{x} \forall S^{\alpha,\beta}, \tag{13}$$

where $S^{\alpha,\beta}$ represents time-space fabric having scaling indices $\alpha$ and $\beta$. The traditional definition of velocity makes no sense in the non-differentiable fractal spacetime. For instance, Feynman[27] observed that the trajectories of quantum mechanical particles are often continuous but nondifferentiable characterized by fractal time-space dimensions[28]. Like the fractional derivative, the Hausdorff derivative exists under a fractal metric spacetime. For instance, $\hat{v} = dt^{1/2}/dt^{1/3}\big|_{t=0}$ exists while $\hat{v} = dt^{1/2}/dt\big|_{t=0}$ does not, and velocity (13) is physically sound for fractal trajectories.

Diffusion processes are governed by the two equations: the continuity equation and the constitutive equation. In terms of the hypotheses of the fractal invariance and equivalence, the former in a fractal $S^{\alpha,\beta}$ is given by

$$\frac{\partial u}{\partial t^\alpha} = -\nabla^\beta \cdot J, \qquad (14)$$

where $\nabla^\beta \cdot$ is the divergence operator on a fractal space, $u$ represents the concentration density of particles, and $J$ denotes particle flux. Likewise, the constitutive Fickian equation on a spatial fractal is stated as

$$J = -D\nabla^\beta u, \qquad (15)$$

where $D$ denotes scale-independent constant diffusivity, and $\nabla^\beta$ is gradient operator on a space having fractal $\beta$. Substituting (15) into (14) produces diffusion equation

$$\frac{\partial u}{\partial t^\alpha} = \nabla^\beta \cdot \left(D\nabla^\beta u\right). \qquad (16)$$

Eq. (16) is actually a time- and space-dependent transport-diffusion equation [see Eq. (A6) further below in Appendix] and can be restated under the fractal time-space fabric $(\hat{t}, \hat{x})$ as a normal diffusion equation,

$$\frac{\partial u}{\partial \hat{t}} = D\hat{\nabla}^2 u, \qquad (17)$$

Where $\hat{\nabla}^2$ is the Laplace operator under coordinate $\hat{x}$. It is easy to see that Eqs. (16) and (17) agree with anomalous diffusion (1) and normal diffusion (5), respectively. Anomalous diffusion equation (16) can be considered a master equation in nature for multidisciplinary applications, where the variable *u* can represent diverse physical quantities, for example, temperature and pore pressure whose corresponding normal diffusion processes (17) involve Fourier's heat conduction law and Darcy's law, respectively.

The Green's function of the Cauchy problem of equation (17) is the time-dependent probability density function (PDF) of normal diffusion under the metric spacetime $(\hat{t}, \hat{x})$. By using transforms (4), this PDF has a stretched Gaussian form under the metric spacetime (*x*, *t*)

$$P(x,t) = \frac{\beta |x|^{\beta-1}}{2(4\pi Dt^\alpha)^{d/2}} e^{-|x|^{2\beta}/4Dt^\alpha} \qquad (18)$$

where *d* is the topological dimensionality. It is known that $\langle \Delta \hat{x}^2 \rangle = 2D\Delta \hat{t}$ in a normal diffusion process and then $\langle \Delta x^{2\beta} \rangle = 2D\Delta t^\alpha$. And the mean square displacement $\sigma^{2\beta} = \langle \Delta x^{2\beta} \rangle = 2D\Delta t^\alpha$ and the PDF (18) is rewritten as

$$P(x,t) = \frac{\beta |x|^{\beta-1}}{2(2\pi \sigma^{2\beta})^{d/2}} e^{-|x|^{2\beta}/2\sigma^{2\beta}}. \qquad (19)$$

In Appendix a comparison is made between the present stretched Gaussian and those reported in the literature. In particular, it is noted that the time-dependent PDF (19) appears somewhat similar to the stretched exponential asymptotics derived from Fox-function PDF arising from anomalous diffusion equation (2) with fractional time derivative[4] ($\beta=1$, $\eta=\alpha$),

$$P_f(x,t) = \frac{|x|^{\mu-1}}{Bt^{\alpha\mu/2}} e^{-b|x|^{\mu}/t^{\alpha\mu/2}}, \qquad (20)$$

where $\mu = 2/2-\alpha$, $B$ and $b$ are coefficients depending on $\mu$, $\eta$ and diffusivity, $0 \prec \mu \prec 2$. Despite the similar mathematical appearance, the stretched Gaussians (18) and asymptotics (20) are different in the definition of the respective time and space exponents.

The spatial Fourier transformed PDF (19) is given by

$$P(k,t) = e^{-Dk^{2\beta}t^{\alpha}}, \qquad (21)$$

which characterizes the relaxation for a fixed wavenumber $k$ and is very similar to the Kohlrausch-Williams-Watts stretched Gaussian[29] and deviates from the classical exponential Debye pattern.[30] The relaxation of anomalous diffusion equation (2) is described by a Mittag-Leffler function of $t^{\alpha}$ which observes transitions from the initial stretched exponential behavior like (21) to a long-time asymptotic inverse power-law behavior $t^{-\alpha}$.[31]

## IV. CONCLUDING REMARKS

It is well known that anomalous diffusion is often associated with a variety of frequency power law scaling phenomena[2-6] mostly involving soft matter such as glass, colloids, emulsions, biomaterials, oil, and various porous media, where the large amount of the elementary molecules is grouped together and behaves like a macromolecule with entangled (non-lattice) and porous mesostructures. The very existence of many-particle long-range interactions and history-dependent motions causes fractal mesoscopic metric spacetime of macromolecules which inflicts a profound impact on various physical behaviors.

In a statistical description or a phenomenological modeling, the fractal has long been considered responsible for anomalous physical behaviors and is claimed to have links with fractional derivatives, Lévy statistics, fractional Brownian motion, and empirical power law scaling.[17] This study made a step forward to present the fractal spacetime transforms and the hypotheses of fractal invariance and equivalence to display explicitly how the fractal metric spacetime influences physical behaviors. Accordingly, the fractional quantum relationships were derived and the fractional Schrodinger equation was found to be a consequence of the fractal spacetime structure. We also introduced the new concept of Hausdorff derivative based on the fractal spacetime transforms and then developed a novel modeling equation for anomalous diffusion, whose Green's solution is new stretched Gaussians.

Although the hypotheses of fractal invariance and equivalence are presented in somewhat heuristic way in this study and need further be solidified in the future research, the present theoretical framework is physically sound and mathematically consistent from anomalous diffusion to statistics and macromechanics to mesoscopic quantum mechanics. Both the traditional fractional derivative and the new Hausdorff derivative are mathematical modeling formalisms underlying the scale spacetime transforms (4). For instance, the inverse of the fractional time transform in (4) is also the kernel function in the definition of the fractional time derivative.[13] However, the fractional derivatives in space and time are non-local, whereas the Hausdorff derivative are local. Both derivatives can give the generalized interpretation of diverse physical concepts on fractal spacetime. On the other hand, the Tsallis distribution has also in recent years been a popular approach in the description of anomalous diffusion. Like the present stretched Gaussians (18) and (19), this distribution was also a solution of the linear varying-coefficient Fokker-Planck equation of transport-diffusion type,[32] in which the standard local integer-order derivatives are used. The corresponding Tsallis nonextensive thermodynamics is claimed capable to describe the long-range interacting systems and memory processes. This shows that the fractional derivative may not be the only approach in modeling anomalous diffusion process. The links and differences between the fractional and the Hausdorff derivatives for fractal spacetime modeling are currently a subject under active study.

**ACKNOWLEDGEMENTS:**

The work described in this paper was partially supported by a grant from the CAEP, China (Project No. 2003Z0603). The author gratefully acknowledges the support of K. C. Wong Education Foundation, Hong Kong.

**APPENDIX**

In literature, some stretched Gaussians are constructed artificially with little links to partial differential equations (PDE) but widely used in diverse situations such as the fitting of anomalous distribution of turbulence experiment data in the form of $C \exp\left(-\frac{|x|^2}{\left[1+(a|x|/\sigma)^v\right]\sigma^2}\right)$, where $C$, $a$ and $v$ are the fitting parameters.[33] A few stretched

Gaussians, however, also underlie a PDE. For instance, the Richardson's turbulent diffusion equation of spherical symmetry is given by[9]

$$\frac{\partial P_R}{\partial t} = \frac{1}{r^{d-1}} \frac{\partial}{\partial r} \left( k_0 r^{d+1-2\beta} \right) \frac{\partial}{\partial r} P_R, \quad (A1)$$

where $k_0$ is a constant coefficient, $r$ represents the radial distance, and $\beta$ is defined as in the text body. The Green's function PDF of equation (A1) is of stretched Gaussian type

$$P_R(r,t) = \frac{\beta \Gamma(d/2)}{\pi^{d/2} \Gamma(1/\beta)(2k_0 \beta t)^{d/2\beta}} e^{-r^{2\beta}/4k_0\beta^2 t}. \quad (A2)$$

In the Richardson case, $\alpha=1$, $\beta=2/3$. The diffusion equation (16) of spherical symmetry under $\alpha=1$ is given by

$$\frac{\partial P_s}{\partial t} = \frac{1}{\hat{r}^{d-1}} \frac{\partial}{\partial \hat{r}} \left( D \hat{r}^{d-1} \right) \frac{\partial}{\partial \hat{r}} P_s = \frac{1}{r^{\beta d-1}} \frac{\partial}{\partial r} \left( D r^{\beta d-2\beta+1} \right) \frac{\partial}{\partial r} P_s. \quad (A3)$$

The present stretched Gaussian PDF corresponding to (A3) is

$$P_s(r,t) = \frac{2\beta r^{\beta+1}}{(4\pi Dt)^{3/2}} e^{-r^{2\beta}/4Dt}. \quad (A4)$$

The difference between Richardson's (A3) and the present PDFs (A4) is evident. Without loss of generality, let us consider the one-dimensional symmetric problem and analyze the difference between these two models. The Richardson equation is expressed as

$$\frac{\partial P_R}{\partial t} = \frac{\partial}{\partial r} \left( k_0 r^{2-2\beta} \right) \frac{\partial P_R}{\partial r}. \quad (A5)$$

In contrast, the present diffusion equation (16) is stated as

$$\frac{\partial P}{\partial t} = -Dr^{1-2\beta} \frac{\partial p}{\partial r} + \frac{\partial}{\partial r} \left( Dr^{2-2\beta} \frac{\partial P}{\partial r} \right). \quad (A6)$$

Obviously, the above equation (A6) is of a transport-diffusion model with time- and space-dependent coefficients, while the Richardson equation (A5) is of a pure diffusion model with a space-dependent diffusivity reflecting the power law scaling.